# An Evolutionary Game based Secure Clustering Protocol with Fuzzy Trust Evaluation and Outlier Detection for Wireless Sensor Networks

Liu Yang, Yinzhi Lu, Simon X. Yang, *Senior Member, IEEE*, Yuanchang Zhong,
Tan Guo, and Zhifang Liang

*Abstract*—Trustworthy and reliable data delivery is a challenging task in Wireless Sensor Networks (WSNs) due to unique characteristics and constraints. To acquire secured data delivery and address the conflict between security and energy, in this paper we present an evolutionary game based secure clustering protocol with fuzzy trust evaluation and outlier detection for WSNs. Firstly, a fuzzy trust evaluation method is presented to transform the transmission evidences into trust values while effectively alleviating the trust uncertainty. And then, a K-Means based outlier detection scheme is proposed to further analyze plenty of trust values obtained via fuzzy trust evaluation or trust recommendation. It can discover the commonalities and differences among sensor nodes while improving the accuracy of outlier detection. Finally, we present an evolutionary game based secure clustering protocol to achieve a trade-off between security assurance and energy saving for sensor nodes when electing for the cluster heads. A sensor node which failed to be the cluster head can securely choose its own head by isolating the suspicious nodes. Simulation results verify that our secure clustering protocol can effectively defend the network against the attacks from internal selfish or compromised nodes. Correspondingly, the timely data transfer rate can be improved significantly.

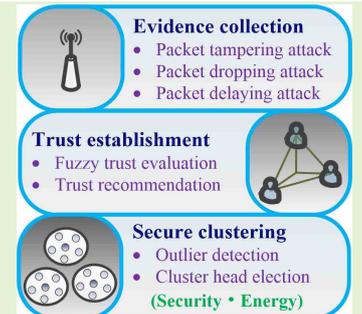

## I. INTRODUCTION

WIRELESS Sensor Networks (WSNs) are based on large numbers of low-cost, low-power, and resource-limited sensor nodes with sensing, processing, and communication abilities [1]. Once deployed into the target area, these sensor nodes can automatically organize into an ad-hoc network to perform sensing tasks. WSNs link the physical world with digital computational systems where these sensed information can be used to achieve real-time decision making [2].

Because of the volatile nature and dynamic topology of WSNs [3] [4], sensor nodes in the network collaborate with one another to execute sensing task and forward data through the network hop by hop based on some specific routing protocols [5]. WSNs are usually deployed in harsh or hostile environment and work in an unsupervised mode, they are vulnerable to various kinds of attacks especially during the routing process [6], such as selective forwarding, Greyhole, Wormhole, and Blackhole attacks [7], [8]. Traditional cryptography-based secure techniques are highly complex and inconsistent with the specific characteristics and severe constraints of WSNs [9]. Moreover, they cannot prevent the attacks from internal selfish or compromised nodes with legitimate identities [10]. To solve these problems, trust management schemes are usually adopted to enhance the reliability and improve the quality of WSNs [11]–[13]. Generally, trust is defined as "confidence in or reliance on some quality or attribute of a person or thing, or the truth in a statement" [14]. A sensor node in the network can cooperate with the reliable adjacent nodes according to the trust management system, and it can also report the monitored suspicious behaviors of neighbors to help to enhance or update the trust management system [15].

### A. Motivation

To establish the trust relationship between sensor nodes, each node usually evaluates trust values of other nodes based on the directly or indirectly monitored transmission behaviors [16]. However, transmission misbehaviors may occur due to

This work was supported in part by the National Natural Science Foundation of China under Grant 61801072, in part by the Chongqing Science and Technology Commission under Grant cstc2018jcyjAX0344, in part by the Science and Technology Research Program of Chongqing Municipal Education Commission under Grant KJQN202000641, and in part by the National Key Research and Development Program of China under Grant 2019YFB2102001. (*Corresponding authors*: *Liu Yang*; *Yuanchang Zhong*).

L. Yang, Y. Lu, T. Guo, and Z. Liang are with the School of Communication and Information Engineering, Chongqing University of Posts and Telecommunications, Chongqing, China (e-mail: yangliu@cqupt.edu.cn; henanluyinzhi@163.com; guot@cqupt.edu.cn; liangzf@cqupt.edu.cn).

S. X. Yang is with the Advanced Robotics and Intelligent Systems Laboratory, School of Engineering, University of Guelph, Guelph, ON N1G2W1, Canada (e-mail: syang@uoguelph.ca).

Y. Zhong is with the State key laboratory of power transmission equipment & system security and new technology, School of Electrical Engineering, Chongqing University, Chongqing 400044, China (e-mail: zyc@cqu.edu.cn).

node failure or congestion, latency, and loss in an open wireless medium [17]. All these issues make the trust evidences acquired by each node inaccurate, incomplete, and imprecise. The uncertainty of trust evidences has negative impact on trust evaluation since a sensor node cannot distinguish the malicious behaviors from false negative ones. The trust level of a normal node can be degraded if several transmission misbehaviors are monitored, and a malicious node can improve its trust level by performing cooperative behaviors. Hence, a single trust value cannot be used directly to detect whether a node is malicious or not. Plenty of trust values obtained by a node should be further analyzed to discover the commonalities and differences among sensor nodes. In addition, energy is a kind of scarce resource for sensor nodes in WSNs. Additional energy cost has to be paid to achieve security assurance so that it is impossible to save energy and assure security simultaneously. Then a trade-off has to be considered to deal with this conflict while enhancing the overall network performance.

### B. Contribution

Based on the above arguments, this paper combines secure clustering, trust evaluation, and outlier detection with a focus on preventing malicious nodes from being cluster heads and achieving a trade-off between energy saving and security assurance in WSNs. Then an evolutionary game based secure clustering protocol with fuzzy trust evaluation and outlier detection (EGSCFO) is proposed. The main contributions of this paper are listed as follows:

1) We present an interval type-2 (IT2) fuzzy logic based trust evaluation method to transform the transmission evidences including data tampering, packet dropping, and transmission delay into trust values. Our method can effectively deal with the trust uncertainty problem while enhancing the accuracy of trust evaluation.

2) A K-Means based outlier detection scheme is proposed to further analyze plenty of trust values acquired by a sensor node. The commonalities and differences among sensor nodes can be discovered so that the malicious nodes are detected with high accuracy.

3) We present an evolutionary game based secure clustering protocol to isolate the malicious nodes from being the cluster heads while a trade-off between security assurance and energy saving can be achieved.

4) Plenty of experiments are performed to evaluate the performance of the proposed secure clustering protocol. Results demonstrate that the proposed protocol outperforms other ones on isolating the malicious nodes from being cluster heads and improving the performance of timely data transfer.

The rest of this paper is organized as follows: The related works are summarized in section II. System model including network model, radio model, and security model is described in section III. Our fuzzy trust evaluation method and outlier detection scheme are presented in sections IV and V respectively. The detail of our secure clustering protocol is proposed in section VI. We verify the performance of our secure clustering protocol in section VII. And finally we conclude this paper in section VIII.

## II. RELATED WORKS

Recently, a substantial amount of researches are performed in WSNs or ad-hoc networks where security problems are critically considered. And many trust and reputation systems are proposed to accurately deal with the secure flat routing or clustering issues.

### A. Trust based secure flat routing

An active detection-based security and trust routing protocol called ActiveTrust is proposed for WSNs [18]. It actively creates a number of detection routes to rapidly detect and obtain nodal trust which is the weighing of direct trust and recommendations. The generation and distribution of detection routes can be created according to the desired security level and energy efficiency. ActiveTrust can effectively enhance the data routing security. To some extent, it can avoid the attacks against the trust management system since a node needs some continuous cooperation to be reconsidered as a router if this node is detected to be malicious recently. However, as one of the most important factors for routing protocol, energy is not considered during the route discovery designing.

A belief based trust evaluation mechanism called BTEM is proposed to isolate the malicious nodes from being the trust-worthy for WSNs [19]. Bayesian estimation is adopted to gather direct and recommendation trust values while the correlation of data collected over the time is further considered. The output probability of a node to be malicious is compared against a pre-defined threshold to make final decision for secured delivery. BTEM can effectively detect malicious nodes while improving the false-positive detection rate. However, it is with low robustness since the final trust value of a node is the fair weighing of direct trust and the recommended one. In addition, the pre-defined threshold is lack of adaptiveness.

To improve the quality of service (QoS) of safety applications in vehicular sensor networks, a trust-based security adaption mechanism (TSAM) is presented [20]. For a vehicular sensor node, the trust level of a link to the neighbor is estimated based on connectivity duration, security level, and centrality metrics of the neighbor. And then the average trust level for all links is calculated as the trust level of this vehicular sensor node. Since the trust level of a vehicle indicates the confidence of this vehicle has on its neighbor nodes, the security level of a vehicle decreases with the increase of its trust level. That is, a vehicle with higher trust level adaptively selects a lower security level for energy-efficient data transmission. TSAM can achieve a trade-off between QoS and security. However, the transmission behaviors are not considered during the trust estimation process.

A trust management-based secure routing scheme (TMSRS) is proposed for industrial sensor networks with fog computing [21]. To update the trust values of sensor nodes, a mathematical Gaussian distribution model is constructed based on the numbers of historical interaction information. And a grey decision making method is adopted to explore the reliable and energy-efficient routing nodes. A trade-off between security, energy, and transmission performance can be achieved in TMSRS. However, continuous noncooperation only results in a slow decline of trust value that may cause attacks against the trust management system.

### B. Trust based secure clustering

Cluster based network model is more suitable for sensor networks than its flat counterpart since it is more energy efficient and scalable.

To adopt trust management system in clustered WSNs, a secure double cluster heads model (DCHM) is presented [15]. It includes cluster, cluster heads, and base station modules. The cluster module considers cluster heads election, clustering, and trust system construction or update. The tasks of outlier detection, credible data fusion, and results uploading are performed by cluster heads. Two cluster heads are selected within a cluster to independently perform the same task. The dissimilarity coefficient of data from the two cluster heads is calculated by base station. A feedback is returned to cluster to help to update trust management system. DCHM can effectively isolate the malicious from the network since two cluster heads independently perform malicious nodes detection within each cluster. However, the communication overhead of double cluster heads is considerable.

A synergetic trust model based on SVM (STMS) is presented for underwater WSNs [22]. The network is divided into some clusters. Each one is composed of several member nodes, a master cluster head (MCH), and a slave head (SCH). Trust evidences including communication, packet, and energy are periodically gathered by member nodes and then reported to the corresponding MCH. K-means algorithm is performed by MCH to group trust evidences into good or bad sets with the label 0 or 1. In addition, SVM algorithm is further performed to classify the labeled evidences into several categories. The results are fed back to member nodes to help to detect the malicious. The trust of MCH is periodically evaluated by the SCH based on the trust evidences from member nodes. If a MCH is detected to be malicious, it has to be replaced by a normal node. Supervised and unsupervised learning algorithms are well combined to establish trust relationship among sensor nodes in STMS. However, extra energy cost has to be paid to select double cluster heads.

To solve the uncertainty problem of trust relationship, a secure clustering scheme with cloud based trust evaluation method (SCCT) for clustered WSNs is proposed [23]. Multi-factors including communication, message, and energy are considered to build trust clouds. These factor clouds are assigned with adjusted weights to calculate immediate trust cloud. The final trust cloud is acquired by synthesizing the recommendation and immediate trust clouds based on time sensitive factor. Then it is converted into trust grade according to its simplicity to the standard grade trust clouds. SCCT can effectively deal with the uncertainty of trust evidences. However, an accurate match between the final trust cloud and standard ones requires much computation cost.

A taylor kernel fuzzy C-means clustering (TKFCC) protocol is proposed to select trust and energy aware cluster heads for sensor networks [24]. Clusters are firstly constructed based on taylor kernel fuzzy C-means algorithm. And then the cluster head is selected within each cluster according to an acceptability factor which is the weighing of fitness constraints including maximal energy, minimal distance, and maximal trust. A well distribution of clusters can be achieved in TKFCC. However, the energy consumption between sensor nodes is not well balanced as the fixed weighing of fitness constraints.

A joint trust based secure routing (JTSR) protocol is proposed for assisting secure communication in WSNs [25]. Firstly, an algorithm named Multi-Objective Taylor Crow Optimization (MOTCO) is adopted to select the optimal cluster heads using multiple objectives including distance, energy, delay, and traffic density. And then, a joint trust, which considering integrity, consistency, forwarding rate, and availability factors, is used to determine whether the obtained optimal cluster heads are trusted or not. Finally, to assure secure delivery to the trusted cluster head for a source node, an algorithm called ChickenDragonfly is adopted to derive the optimal routing path. JTSR can effectively improve the security performance of the network. However, the joint trust is calculated in a centralized mode that is lack of scalability.

A secured QoS aware energy efficient routing (SQEER) protocol is designed to enhance network security while optimizing energy utilization for WSNs [26]. Firstly, both direct and indirect trust values are calculated while spatial and temporal constraints are included to construct the trust relationship between sensor nodes. And then, cluster heads are selected according to trust scores and QoS metrics to perform cluster based secure routing. Finally, path-trust, energy, and hop count are considered to acquire the final routing path. SQEER can well make QoS assurance. However, the recommended maximum trust value from trusted nodes is used as the final recommendation that may be inappropriate.

A fuzzy temporal rule and cluster-based secured routing with outlier detection (FRCSROD) algorithm is proposed for secure communication in WSNs [27]. FRCSROD performs a temporal reasoning-based clustering method to construct dynamic clusters according to distance and residual energy. The nodes with high energy, low mobility, and high trust are selected as cluster heads. In addition, to isolate the malicious nodes from clusters and communication, a fuzzy rule and distance-based outlier detection method is presented which incorporates a security model consisting of a trust mechanism and an authentication process. FRCSROD can effectively improve the reliability of communication, packet delivery ratio, end-to-end delay, and energy consumption. However, the authentication process performed by each node requires much communication and computation overhead.

## III. SYSTEM MODEL

In this section, we give the system model which includes network model, radio model, and security model.

### A. Network model

Hierarchical network model is considered in this paper since it is more flexible and efficient for WSNs [28], as shown in Fig. 1. Sensor nodes in the network are divided into some groups called clusters. And each one consists of a cluster head and some member nodes. All member nodes located in the bottom layer perform the task of data sensing for interest. Cluster heads in the middle layer construct the routing backbone to perform tasks of gathering, integrating, and forwarding the data from member nodes. The base station in the top layer acts as the relay node for the data from cluster heads to the server.

### B. Radio model

Due to the interferences in an open wireless medium, the state of channel quality is unstable that may be good or bad occasionally. Then a Markov chain with two states $S = \{s_0, s_1\}$ is adopted in this paper to model the time-varying wireless medium [29], as shown in Fig. 2. Here $s_0$ and $s_1$ represent the bad and good states of the channel quality respectively. The time interval $t$ that the channel quality becomes each state is a random variable which follows an exponential distribution as

$$p(t) = \begin{cases} \alpha_i e^{-\alpha_i t} & t \geq 0 \\ 0 & t < 0 \end{cases}, \quad (1)$$

where $\alpha_i$, $i \in \{0, 1\}$ are the rates of bad and good states. Then the probabilities of the channel sate to become bad and good are calculated by $p_0 = \alpha_0/(\alpha_0 + \alpha_1)$ and $p_1 = \alpha_1/(\alpha_0 + \alpha_1)$ respectively.

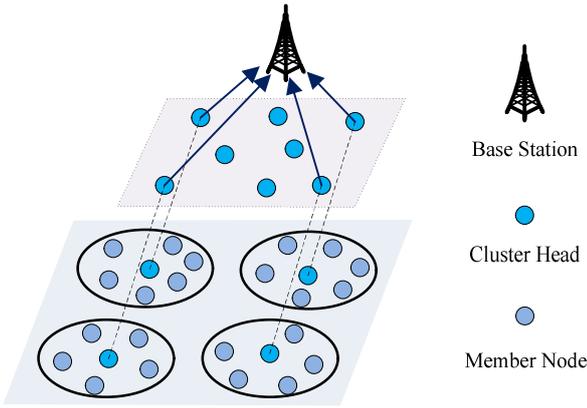

Fig. 1. Cluster based network model.

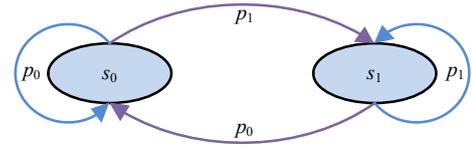

Fig. 2. Markov chain based channel quality model.

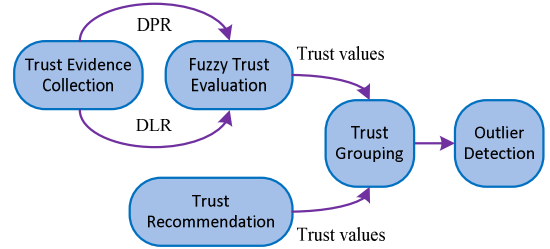

Fig. 3. Our trust based security model.

The wireless transmission loss can either be described by free-space propagation model or two-ray ground reflection model based on the distance $d$ from transmitter to receiver [28]. If $d$ is smaller than a threshold $d_0$, then the first model is more suitable to reflect the transmission loss. Otherwise, the second model is acceptable. Here the threshold $d_0$ can be computed by

$$d_0 = \sqrt{\varepsilon_{fs}/\varepsilon_{amp}}, \quad (2)$$

where $\varepsilon_{fs}$ and $\varepsilon_{amp}$ are the amplified characteristic constants regarding the transmission loss models.

The energy [1] [28] spent by a node for transmitting a $k$-bit data packet to the destination with distance $d$ can be denoted as

$$E_{Tx}(k, d) = \begin{cases} kE_{elec} + k\varepsilon_{fs}d^2, & d < d_0 \\ kE_{elec} + k\varepsilon_{amp}d^4, & d \geq d_0 \end{cases}, \quad (3)$$

where $E_{elec}$ is the energy spent by the transmitter or receiver circuitry.

The energy consumed by the receiver for a $k$-bit data packet can be computed by

$$E_{Rx}(k) = kE_{elec} + kE_{DA}, \quad (4)$$

where $E_{DA}$ is the energy consumed by the receiver for aggregating a one-bit packet.

### C. Security model

To prevent the attacks in resource-constrained WSNs, this paper presents a trust based security model which consists of trust evidence collection, fuzzy trust evaluation, trust recommendation, trust grouping, and outlier detection, as shown in Fig. 3. To establish the trust based secure system, each sensor node periodically collects the trust evidences of other nodes by overhearing the transmissions. Once the evidences are updated, an IT2 fuzzy logic system (FLS) is introduced to estimate the trust values via fuzzy inferring that can effectively alleviate the uncertainty of trust evidences. Combined with trust recommendation mechanism, each node can acquire plenty of trust values which are further analyzed by trust grouping. Then the group means are used to detect the outliers while whether a node is malicious or not can be identified.

Since this paper aims at assuring the security of packet transmissions, trust evidences including packet dropping rate (DPR) and delaying rate (DLR) are collected that can indicate the changes regarding packets when attacks occur during the routing process. Generally, three kinds of attacks regarding data packets can be launched by malicious nodes: tampering, dropping, and delaying. To prevent against tampering attack, data packets need to be authenticated by the destination node to ensure integrity. Any packet which is found to be tampered should be discarded due to authentication failure, which implies that the tampering attack can be somehow regarded as dropping attack from the perspective of sender nodes [30]. Hence, the two trust evidences DPR and DLR are able to indicate all kinds of attacks against data packet to some degree.

## IV. FUZZY TRUST EVALUATION

IT2 FLS has been proved better ability to deal with uncertainty and smoother control surface than its type-1 (T1) counterpart [31]. The major difference is that at least one of the fuzzy sets in the rule base is an IT2 fuzzy set. Fig. 4 gives the schematic diagram of an IT2 FLS [32]. Firstly, the fuzzifier module normalizes the crisp inputs into fuzzy sets. And then, the inference engine infers a solution according to the input fuzzy sets and the rule base which defines the relationships between inputs and output with *IF THEN* clauses. Thirdly, type reduction is performed by the type reducer while a T1 fuzzy set is acquired. Finally, the defuzzifier gives crisp output via denormalization.

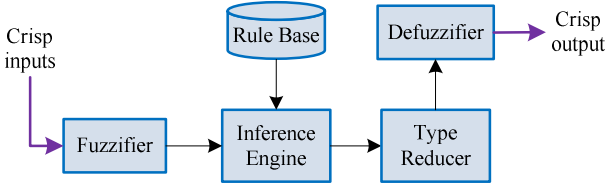

Fig. 4. Schematic diagram of a typical IT2 FLS.

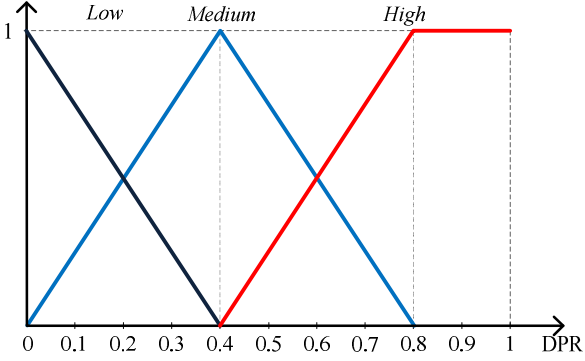

Fig. 5. Fuzzy sets designed for DPR.

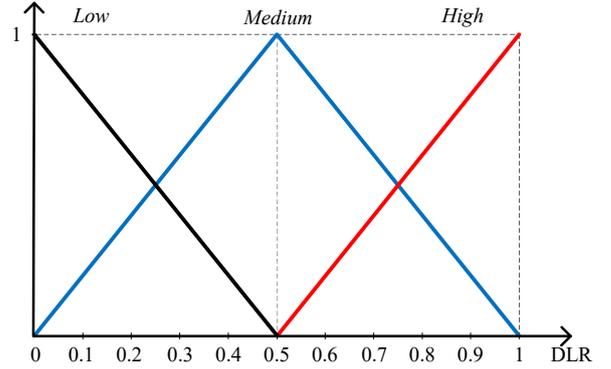

Fig. 6. Fuzzy sets designed for DLR.

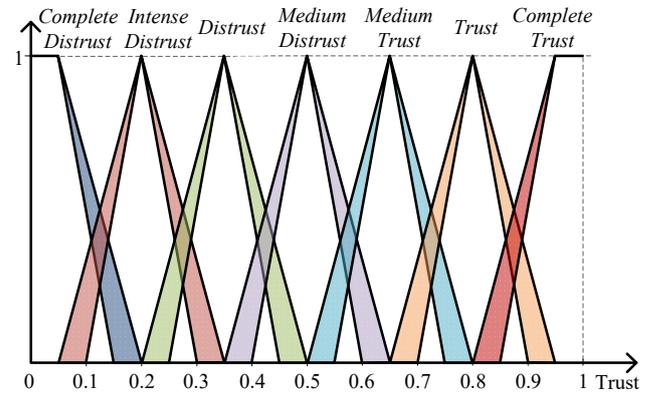

Fig. 7. IT2 fuzzy sets designed for the trust output.

To evaluate the trust of sensor nodes, the acquired trust evidences DPR and DLR are used as the inputs of the FLS. Three fuzzy sets corresponding to the linguistic variables *low*, *medium*, and *high* are adopted to describe the membership grade of DPR, as shown in Fig. 5. Similarly, the fuzzy sets for DLR are shown in Fig. 6. Seven IT2 fuzzy sets corresponding to the variables *complete distrust*, *intense distrust*, *distrust*, *medium distrust*, *medium trust*, *trust*, and *complete trust* are designed to assess the membership grade of the trust output, as shown in Fig. 7. To define the mapping of fuzzy sets between the trust evidences and trust output, nine fuzzy rules are established with *IF THEN* clauses, as shown in Table I. Such as the first rule: If DLR is *low* and DPR is *low*, then the trust output is *complete trust*.

Let $x = (x_1, x_2)$ be the input vector of the FLS. Here $x_1$ and $x_2$ represent the input variables regarding DPR and DLR respectively. To acquire the output trust value, seven steps are performed as follows:

1) Calculate the membership grade for $x_i$ ($i = 1, 2$) based on the corresponding fuzzy set of the $k^{th}$ ($k = 1, 2, ..., 9$) rule, expressed as $G(x_i^k)$.

2) Compute the firing grade $G(x^k)$ regarding the $k^{th}$ ($k = 1, 2, ..., 9$) rule for $x$ according to product $t$-norm [32].

3) Calculate output trust interval $[T_l(x^k), T_r(x^k)]$ of the $k^{th}$ ($k = 1, 2, ..., 9$) rule for $x$. Here $T_l$ and $T_r$ are the left and right trust values respectively. Especially, the trust fuzzy set of the $k^{th}$ ($k = 2, 3, ..., 8$) rule is vertically symmetrical, then two output trust intervals corresponding to the left and right part are acquired. Since the two intervals share the same firing grade, the grade of each one should be halved. Then we can get some value pairs of trust interval versus grade that can be represented as a trust set $\{<[T_l(x^t), T_r(x^t)], G(x^t)>, t = 1, 2, ...,16\}$.

4) Covert the value pairs of trust interval versus grade into the ones of trust value versus grade interval, then the trust set can be reconstructed as $\{<T_a(x^t), [G_l(x^t), G_u(x^t)]>, t = 1, 2, ...,16\}$. Here $T_a$ is the average of the corresponding left and right trust values. $G_l$ and $G_u$ are the lower and upper membership grades of $T_a$ regarding to the corresponding trust fuzzy set.

5) Normalize all lower membership grades and upper ones in the trust set separately while sorting all trust values in

TABLE I
FUZZY RULES OF TRUST EVALUATION

| No. | DLR | DPR | Trust output |
|---|---|---|---|
| 1 | *Low* | *Low* | *Complete Trust* |
| 2 | *Medium* | *Low* | *Trust* |
| 3 | *High* | *Low* | *Medium Trust* |
| 4 | *Low* | *Medium* | *Medium Trust* |
| 5 | *Medium* | *Medium* | *Medium Distrust* |
| 6 | *High* | *Medium* | *Distrust* |
| 7 | *Low* | *High* | *Distrust* |
| 8 | *Medium* | *High* | *Intense Distrust* |
| 9 | *High* | *High* | *Complete Distrust* |

ascending order. Then the trust set can be updated as $\{<T_a(x^v), [G_l(x^v), G_u(x^v)]>, v = 1, 2, ...,16\}$.

6) Use center-of-sets [33] method to perform type reduction based on the trust set. Then a T1 fuzzy set $1/[T_L(x), T_R(x)]$ for the input vector $x$ can be obtained. Here $T_L$ and $T_R$ are the final left and right trust values that can be calculated by

$$T_L(x) = \min_{i=1,2,...,15} \frac{\sum_{v=1}^{i} G_u(x^v) \cdot T_a(x^v) + \sum_{v=i+1}^{16} G_l(x^v) \cdot T_a(x^v)}{\sum_{v=1}^{i} G_u(x^v) + \sum_{v=i+1}^{16} G_l(x^v)}$$
$$= \frac{\sum_{v=1}^{I} G_u(x^v) \cdot T_a(x^v) + \sum_{v=I+1}^{16} G_l(x^v) \cdot T_a(x^v)}{\sum_{v=1}^{I} G_u(x^v) + \sum_{v=I+1}^{16} G_l(x^v)}, \quad (5)$$

$$T_R(x) = \min_{j=1,2,...,15} \frac{\sum_{v=1}^{j} G_l(x^v) \cdot T_a(x^v) + \sum_{v=j+1}^{16} G_u(x^v) \cdot T_a(x^v)}{\sum_{v=1}^{j} G_l(x^v) + \sum_{v=j+1}^{16} G_u(x^v)}$$
$$= \frac{\sum_{v=1}^{J} G_l(x^v) \cdot T_a(x^v) + \sum_{v=J+1}^{16} G_u(x^v) \cdot T_a(x^v)}{\sum_{v=1}^{J} G_l(x^v) + \sum_{v=J+1}^{16} G_u(x^v)}, \quad (6)$$

where $I$ is left switch point [33] determined by $T_a^I < T_L < T_a^{I+1}$. $T_a^I, T_a^{I+1} \in \{T_a^v, v = 1, 2, ...,16\}$. $J$ is right switch point follows $T_a^J < T_R < T_a^{J+1}$. $T_a^J, T_a^{J+1} \in \{T_a^v, v = 1, 2, ...,16\}$. $T_L$ and $T_R$ can be acquired by using algorithm EIASC [32].

7) The final output trust value $T(x)$ for the input $x$ can be acquired through defuzzification:

$$T(x) = \frac{T_L(x) + T_R(x)}{2}. \quad (7)$$

## V. OUTLIER DETECTION

A sensor node, no matter is malicious or not, can enhance its trust level through cooperation. However, its trust level may be degraded due to malicious behaviors or error monitoring in an open wireless medium. Hence, a single trust value cannot be used to accurately detect whether a node is malicious or not. To solve this problem, plenty of trust values acquired by a node should be further analyzed through an outlier detection scheme.

A K-Means based outlier detection scheme is designed in this section which is independently performed by each node to detect the malicious. To accelerate the convergence speed, trust recommendation mechanism is introduced to update the trust values efficiently. Usually, message exchange only happens between the cluster head and member node in clustered WSNs. Then if a sensor node joins in a cluster where the head is trusted, it can request the trust recommendation from this head.

If a sensor node $i$ has received the recommended trust value $T(k, j)$ from cluster head $k$ for node $j$, it updates the trust value of node $j$ as

$$T(i,j) = \begin{cases} \dfrac{T(i,j) + T(i,k)T(k,j)}{1+T(i,k)}, & \text{if } T(i,j) > 0 \\ T(i,k)T(k,j), & \text{otherwise} \end{cases}, \quad (8)$$

**Algorithm I** Outlier detection

| | |
|---|---|
| Initialize | convergence state ← *false*<br>**if** *i* is a cluster member node and length(*TS*) > 1<br>    Randomly select two values $T_a$, $T_b$ from *TS*;<br>    **if** $T_a > T_b$<br>        avHTG ← $T_a$;<br>        avLTG ← $T_b$;<br>    **else**<br>        avHTG ← $T_b$;<br>        avLTG ← $T_a$;<br>    **end**<br>**end** |
| Iterate | **if** *i* is a cluster member node<br>    HTG ← [];<br>    LTG ← [];<br>    Update *TS* if a new round comes;<br>    **for** *k*=1:1:length(*TS*)<br>        **if** \|TS(k)-avHTG\| < \|TS(k)-avLTG\|<br>            HTG ← [HTG TS(k)];<br>        **else**<br>            LTG ← [LTG TS(k)];<br>        **end**<br>    **end**<br>    **if** \|mean(HTG)-avHTG\|, \|mean(LTG)-avLTG\| < $d_m$<br>        break iterate;<br>    **else**<br>        avHTG ← mean(HTG);<br>        avLTG ← mean(LTG);<br>    **end**<br>**end** |
| Converge | $c_1$: \|avHTG - last_avHTG\|, \|avLTG - last_avLTG\| < $d_m$<br>$c_2$: avHTG - avLTG > $d_{mbg}$<br>$c_3$: the times that $c_1$ and $c_2$ are continuously satisfied > $T_s$<br>**if** $c_1$, $c_2$, and $c_3$ are *true*<br>    convergence state ← *true*;<br>**end** |

where $T(i, k)$ is the trust value of node $i$ to $k$ that updated through fuzzy inference. $T(k, j)$ is the trust of node $k$ to $j$ that is acquired via either trust recommendation or fuzzy inference.

The outlier detection process works based on rounds. In each round, a sensor node firstly updates the trust values of other nodes through fuzzy inference or trust recommendation. And then these updated trust values are further analyzed by using K-Means algorithm. The results are finally used to isolate the malicious nodes from being the cluster heads. Initially, no interaction happens among nodes so that each node has to record the trust value as 0 for other nodes. A node does not initiate its own outlier detection process unless it has interacted with more than two nodes. Once the outlier detection process is activated by a node, it continues only in these rounds where this node acts as a cluster member node that has the chance to update the trust values.

The detail of outlier detection process for any node $i$ is given in Algorithm I. Two trust groups are constructed from the trust set (*TS*) which contains all positive trust values. One is called as High Trust Group (*HTG*), and the other is called as Low Trust Group (*LTG*). In each round, several iterations are included to assign the higher and lower trust values into the groups *HTG* and *LTG* respectively according to the K-Means algorithm. The iterations within a round end if the absolute difference between the last two group means is less than a threshold $d_m$ for both trust groups. The final group means are used as the initial ones of the two trust groups in the next round. With the proceeding of the outlier detection process, the two group means tend to be stable due to sufficient trust information. We give the convergence conditions as follows:

1) The absolute difference of final group means between two contiguous rounds is less than the threshold $d_m$ for both trust groups.

2) The difference of group means between *HTG* and *LTG* is more than a threshold $d_{mbg}$.

3) The above two conditions keep satisfied for $T_s$ times.

If the first condition is true, then the means of both trust groups tend to be steady. An obvious difference between the two trust groups exists if the above second condition is satisfied. To avoid the false positive convergence of the outlier detection process, the threshold $T_s$ in the last condition should be much bigger than 1.

## VI. Detail of Our Secure Clustering

### A. Evolutionary Game based cluster head election

Sensor nodes cannot achieve energy saving and security assurance simultaneously when deciding whether to be cluster heads in WSNs. On the one hand, each node expects one of its trusted neighbors to serve as the cluster head so that it can save more energy to acquire longer lifetime. But on the other hand, a node has to campaign for the cluster head to prevent the malicious one from being the head. Since game theory is an effective method to make decisions for selfish and rational individuals under conflict situations [34], in this section we use it to construct the cluster head election model while achieving a trade-off between energy and security.

For any sensor node, it decides whether to be the cluster head by playing a cluster head election game with its past cluster heads which are now detected to be non-malicious. Here we formally define this game as $CEG = \{N, S, U\}$. Where $N$ is the set of game players, $S = \{S_i \mid i \in N\}$ is the strategy combination of all players, and $U = \{U_i \mid i \in N\}$ is the set of utility functions. The payoff matrix for a simple game with two players is shown in Table II. If a player does not declare to be the cluster head while no other player becomes the head, then a malicious node has the chance to be the head while launching attacks that results in a loss $z$ to the entire population. If at least one other player declares to be the cluster head, then it can gain $v$ by enjoying the service of secure data transmission. If a player declares to be the cluster head, then its payoff is $v - c$ due to an additional cost $c$ has to be paid to serve for other nodes. According to the payoff matrix, the utility function $U_i$ for an arbitrary player $i$ can be expressed as

$$U_i = \begin{cases} v, & \text{if } s_i = ND \text{ and } \exists j \in N \text{ s.t. } s_j = D \\ v-c, & \text{if } s_i = D \\ -z, & \text{if } s_j = ND, \forall j \in N \end{cases}. \quad (9)$$

The expected payoff $E(D)$ for a player to declare to be the cluster head can be calculated by

$$E(D) = v - c. \quad (10)$$

TABLE II
PAYOFF MATRIX

|    | D          | ND       |
|----|------------|----------|
| D  | (v-c, v-c) | (v-c, v) |
| ND | (v, v-c)   | (-z, -z) |

The expected payoff $E(ND)$ of not declaring to be the cluster head can be computed by

$$E(ND) = \left(1-(1-p)^{|N|-1}\right)v - (1-p)^{|N|-1}z, \quad (11)$$

where $p$ is mixed strategy that denotes the probability to be the cluster head for a node.

The expected payoff $E_N$ for the entire population $N$ can be expressed as

$$E_N = pE(D) + (1-p)E(ND). \quad (12)$$

We can get the time derivative of the proportion of the population that declaring to be the cluster head, which is shown as

$$\begin{aligned}\frac{dp}{dt} &= p(E(D) - E_N) \\ &= p(1-p)\left((1-p)^{|N|-1}(v+z) - c\right).\end{aligned} \quad (13)$$

The above equation is also called Replicator Dynamics Equation [35], [36]. It reflects the fact that the proportion of the players which declaring to be cluster heads increases (decreases) at time $t$ if their payoffs are larger (smaller) than the expected one of the entire population. The derivative vanishes at three points, expressed as $p_0$, $p_1$, and $p_2$. The first two are $p_0 = 0$, and $p_1 = 1$. And the remainder is denoted by

$$p_2 = 1 - \left(\frac{c}{v+z}\right)^{\frac{1}{|N|-1}}. \quad (14)$$

A point which is larger (smaller) than $p_2$ can acquire a negative (positive) increase rate regarding the proportion of the players which declaring to be cluster heads. That is, $p_2$ is a stable equilibrium point of the game *CEG*, and it is also called as the Evolutionary Stable Strategy (*ESS*) point [35]. In addition, the stability analysis can verify that $p_0$ and $p_1$ are not the *ESS* points.

Let $E_{CH}$ and $E_{CM}$ be the energy spent for data transmission by a cluster head node and member node respectively. The extra cost $c$ for serving as the cluster head can be expressed as the loss of lifetime that formulated by

$$c = \frac{E_{CH} - E_{CM}}{E_{CM}}. \quad (15)$$

The payoff $v$ can be expressed as the desired lifetime with energy $E_{CH}$. It is calculated by

$$v = \frac{E_{CH}}{E_{CM}}. \quad (16)$$

If all players do not declare to be the cluster head, then the loss $z$ can be expressed as the loss of lifetime of the entire population that is calculated by

$$z = |N|(1 - T_{avr}), \quad (17)$$

where $|N|$ is the total number of players, and $T_{avr}$ is the average trust value of the adjacent nodes which are detected to be malicious.

Based on Eqs. (14-17), the *ESS* point of the game *CEG* can be recalculated by

$$p_2 = 1 - \left(\frac{w-1}{w + |N|(1-T_{avr})}\right)^{\frac{1}{|N|-1}}, \quad (18)$$

where $w$ is the multiples of energy spent for data transmission between the cluster head node and member node.

According to the *ESS* of the game *CEG*, a sensor node with more adjacent trusted nodes or higher trust on adjacent malicious ones has less expectation to be the cluster head. In addition, with the increase of the ratio $w$, the expectation to be the cluster head decreases for a node.

### B. Detail of the secure clustering protocol

In our proposed secure clustering protocol EGSCFO, if a sensor node has not been a cluster head during the latest $1/p_{CH}$ rounds, then it self-decides whether to be the cluster head according to a threshold $Th_{CH}$ that calculated by

$$Th_{CH} = p_{CH} / (1 - p_{CH}(r \bmod (1/p_{CH}))), \quad (19)$$

where $p_{CH}$ is the probability to be the cluster head, and $r$ is the current round.

Initially, the probability $p_{CH}$ to be the cluster head for each node is set to be $p_{int}$. With the operating of the network, each node updates its own probability to be cluster head according to Eq. (18). If a sensor node declares to be the cluster head in the current round, then it broadcasts the election message within the network. To choose an appropriate cluster head and then join in the corresponding cluster, a non-cluster head node considers the nearest $N_{NCH}$ cluster heads as the candidates. Before the convergence of the outlier detection process, this non-cluster head node gives priority to the nearer candidate whose trust value is zero. If the trust values of all candidates are nonzero, this non-cluster head node chooses the most reliable candidate as its own head. Once the outlier detection process converges in one round, a non-cluster head node chooses the nearer trusted candidate whose trust value can be grouped into *HTG* to join in cluster while requesting trust recommendations. If no trusted candidate cluster head exists, this non-cluster head node can choose the one whose trust value is zero to be its own cluster head. A sensor node which has not found an appropriate cluster head finally has to self-declare to be the head if it is eligible in the current round.

After cluster structures are constructed in the network, each cluster member node transmits its own data to the cluster head within an allocated time slot. Once a normal cluster head node has received the data from its member nodes, it forwards the data packet to base station immediately after data fusion. However, a malicious cluster head node may forward data packet cooperatively, or tamper data, drop packet, and forward with a random delay. Hence, all transmission behaviors of cluster heads are monitored by their member nodes for fuzzy trust evaluation. And due to the interference in an open wireless medium, the monitored behaviors may not precisely reflect the intentions of the corresponding cluster heads. After a member node updates the trust values of other nodes through fuzzy trust evaluation or trust recommendation, it further performs outlier detection to isolate the malicious nodes from being the cluster heads in the next round.

## VII. PERFORMANCE EVALUATION

### A. Experimental setup

We assume each malicious cluster head launches packet dropping and delaying attacks with the probabilities $P_{DP}$ and $P_{DL}$ respectively. A malicious delay is randomly selected no longer than the maximum duration $D_m$. In addition, we assume that a normal cluster head always forwards the packet without delay for member nodes. And the transmission behaviors can be exactly monitored if the state of channel quality is good. However, the state of channel quality becomes bad if interference occurs, then a normal transmission may be not overheard with the probability $P_{LOS}$. If a normal cluster head does not receive a response from the receiver after transmitted the packet, it has to retransmit the packet with an interval of $0.5 \times D_m$. And this normal retransmission occurs with the probability $P_{DEL}$ if the state of channel quality becomes bad.

Let $E_m$ be the energy spent by a node for keeping monitoring. If a packet is successfully overheard within a duration $d_m$ ($d_m \leq D_m$), an additional energy $E_h$ has to be consumed. Let $E_t$ be the total energy for monitoring a $k$-bit data packet, it can be calculated by

$$E_t(d_m, k) = \begin{cases} d_m E_m + k E_h, & \text{if packet is overheard} \\ D_m E_m, & \text{otherwise} \end{cases}. \quad (20)$$

To verify the performance of our protocol EGSCFO, we compare it with that of SCCT and TKFCC on the MATLAB platform, since trust uncertainty problem is addressed through cloud model in SCCT and a trade-off between energy and security is considered in TKFCC. To perform the comparison experiments, some parameters in SCCT and TKFCC are set as follows: In SCCT, DPR and DLR are used to calculate the trust evaluation factors. The failure tolerance and time sensitive factors are set to be 0.2 and 0.6 respectively. In addition, totally 20 drops are selected randomly from the final trust cloud to perform trust decision-making. In TKFCC, the number of clusters is set to be 5 percent of the total number of sensor nodes. Other parameters used throughout the experiments are listed in Table III.

TABLE III
PARAMETERS SETTING

| Parameter | value | Parameter | value |
|---|---|---|---|
| Initial probability $p_{int}$ to be cluster heads | 0.07 | Maximal overhearing duration $D_m$ (s) | 10 |
| Initial energy $E_0$ (J) | 2 | $\alpha_0$ | 3 |
| Packet size (bits) | 3000 | $\alpha_1$ | 7 |
| Control packet size (bits) | 300 | $d_m$ | 0.05 |
| $E_{elec}$ (nJ/bit) | 50 | $d_{mbg}$ | 0.1 |
| $\varepsilon_{amp}$ (pJ/bit/m$^4$) | 0.0013 | $T_s$ | 60 |
| $\varepsilon_{fs}$ (pJ/bit/m$^2$) | 10 | $w$ | 6 |
| $E_{DA}$ (nJ/bit/message) | 5 | $P_{DP}$ | 0.2 |
| $E_h$ (nJ/bit) | 5 | $P_{DL}$ | 0.2 |
| $E_m$ (nJ/s) | 10 | $P_{LOS}$ | 0.2 |
| $N_{NCH}$ | 2 | $P_{DEL}$ | 0.2 |

### B. Experiment results

In our experiments, we first evaluate the performance of our protocol EGSCFO for the network with 100 sensor nodes randomly deployed into the 100 × 100 m$^2$ sensing area while the base station is placed on the east outside the sensing area. And then, to verify the scalability of our protocol, experiments are further performed for the networks with different sizes and node densities.

Fig. 8 gives the details of average number of malicious clusters per cycle that the malicious nodes are selected as the cluster heads. Here a cycle represents totally 50 rounds of data transmission, and the percentage of the malicious is set to be 20. This figure shows that our protocol EGSCFO has the fewest malicious clusters in most cycles while the average number decreases with fluctuations and rapidly reaches the minimal value 0 with the increase of the cycle. This is because on the one hand, our fuzzy trust evaluation method can deal with the trust uncertainty problem while the accuracy of trust evaluation can be improved. On the other hand, plenty of trust values obtained by a node are further analyzed using our outlier detection scheme, so that the malicious nodes can be effectively isolated from being the cluster heads. Regarding the curve of average number of malicious clusters, this figure shows that EGSCFO terminates firstly, followed by SCCT, and the final is TKFCC. This is because the round where the last non-malicious node dies due to energy exhaustion arrives in this order for the three protocols.

The comparisons of total number of dropping and delaying attacks among EGSCFO, TKFCC, and SCCT are given in Figs. 9 and 10 respectively. These figures show that our protocol EGSCFO has the fewest dropping and delaying attacks among these protocols for all the cases of the percentage of malicious nodes. In addition, with the increase of the percentage of the malicious, the number of attacks increases slightly in EGSCFO while an obvious increase exists in TKFCC and SCCT. Since cloud model is adopted to deal with the uncertainty of trust evidences in SCCT, these figures show that SCCT has fewer attacks than TKFCC for all the cases of the percentage of the malicious.

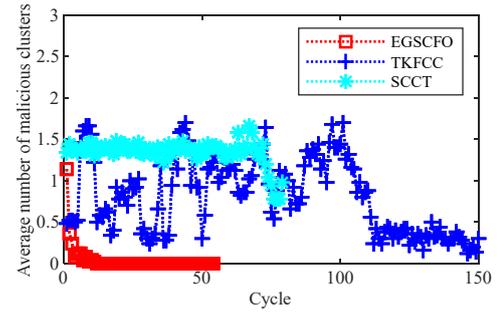

Fig. 8. Average number of malicious clusters per cycle.

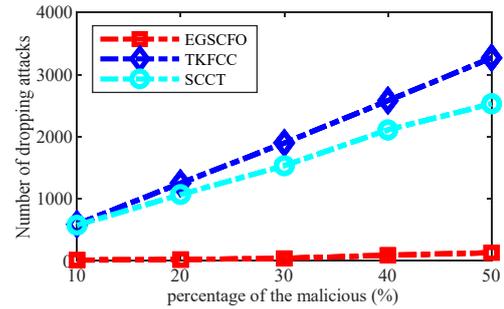

Fig. 9. The comparison of number of dropping attacks among EGSCFO, TKFCC, and SCCT.

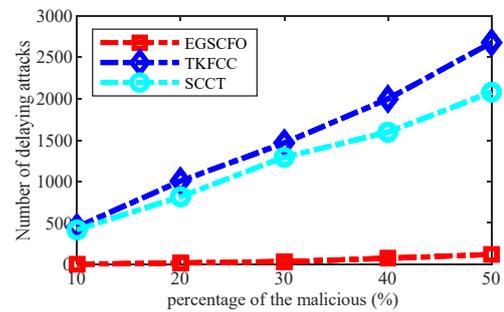

Fig. 10. The comparison of number of delaying attacks among EGSCFO, TKFCC, and SCCT.

Fig. 11 gives the comparison of network lifetime among EGSCFO, TKFCC, and SCCT. Here network lifetime is defined as the total alive rounds of the first dead non-malicious node due to exhaustion of energy. This figure shows that our protocol EGSCFO has shorter network lifetime than SCCT but longer one than TKFCC. This is because more malicious nodes are isolated from being the cluster heads in EGSCFO, so that the normal ones have to burden heavier data transmission tasks. Although more malicious nodes have the chance to act as the cluster heads in TKFCC, the fixed weighing of fitness constraints results in the unbalanced energy consumption among sensor nodes. This figure also shows that with the increase of the percentage of malicious nodes, the network lifetime decreases for all these protocols. Particularly, this case is more obvious in EGSCFO. This is because with the increase of the number of the malicious, the normal nodes have to perform more data transmission tasks. Moreover, the malicious sensor nodes are isolated from being the cluster heads more rapidly in EGSCFO.

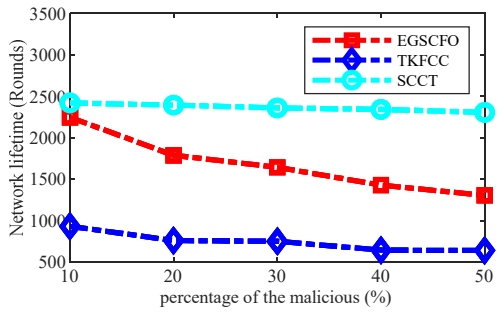

Fig. 11. The comparison of network lifetime among EGSCFO, TKFCC, and SCCT.

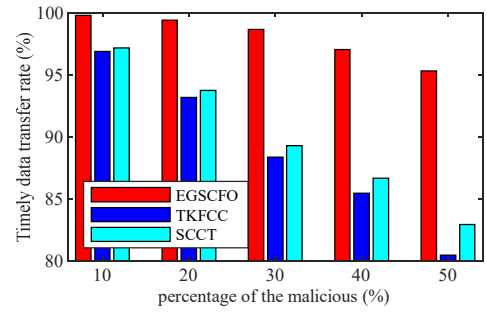

Fig. 13. The comparison of timely data transfer rate among EGSCFO, TKFCC, and SCCT.

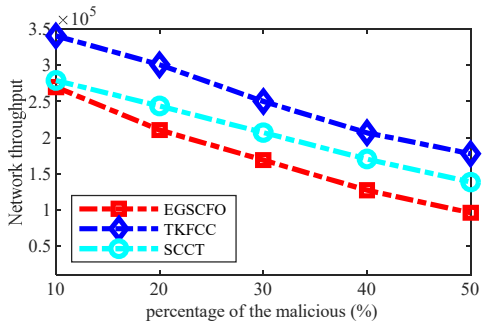

Fig. 12. The comparison of network throughput among EGSCFO, TKFCC, and SCCT.

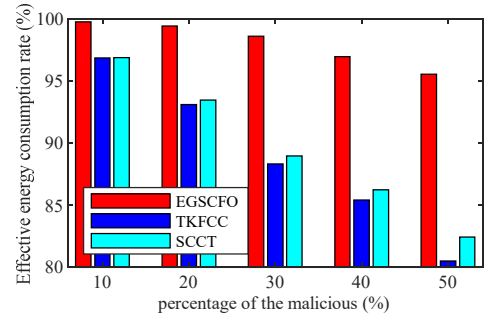

Fig. 14. The comparison of effective energy consumption rate among EGSCFO, TKFCC, and SCCT.

The comparison of network throughput among EGSCFO, TKFCC, and SCCT is given in Fig. 12. Here network throughput is defined as the total number of packets which are collected by non-malicious nodes and successfully transmitted to base station. This figure shows that EGSCFO has the lowest network throughput while TKFCC has the highest. This is because the fewest malicious nodes in EGSCFO have the chance to forward data for non-malicious nodes while in TKFCC the most malicious nodes could act as cluster heads so that the non-malicious ones can save more energy for data collecting. Note that, to enhance the trust level, the malicious nodes are rational enough to cooperatively transmit data for other nodes most of the time.

The comparisons of timely data transfer and effective energy consumption rates among EGSCFO, TKFCC, and SCCT are given in Figs. 13 and 14 respectively. These figures show that EGSCFO has the highest rates of both timely data transfer and effective energy consumption while TKFCC has the lowest. This is because non-cluster head nodes in EGSCFO can join in non-malicious clusters with the highest probability so that they have the least chance to waste energy due to their data packets are most probably transmitted to base station in time. Whereas, malicious nodes in TKFCC have the most chance to act as cluster heads while launching data dropping or delaying attacks that results in the waste of energy for non-cluster head nodes. These figures also show that with the increase of the percentage of malicious nodes, the rates of timely data transfer and effective energy consumption decrease for these protocols. This is because more attacks may occur during the routing process for the network with more malicious nodes.

To verify the scalability of our protocol, the following experiments are performed for the networks with different sizes and node densities. And the percentage of the malicious is set to be 30.

Fig. 15 gives the comparison of total number of attacks including packet dropping and delaying for the networks with different sizes and node densities. This figure verifies that our protocol EGSCFO has the fewest attacks for all the scenarios that owes to our trust based secure mechanism which includes fuzzy trust evaluation and K-Means based outlier detection. For the networks with the same size, this figure shows that the one with bigger node density suffers more attacks for all the protocols. This is because more malicious nodes exist in the network with bigger node density. For the networks deployed with the same number of sensor nodes, this figure also shows that the one with smaller size suffers more attacks. This is

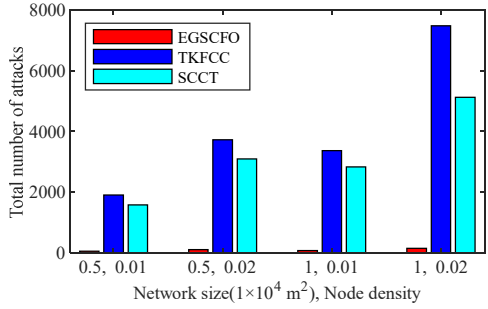

Fig. 15. The comparison of total number of attacks for the networks with different sizes and node densities.

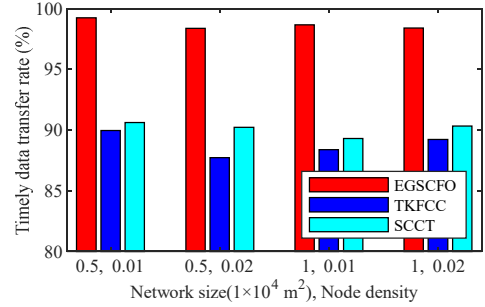

Fig. 17. The comparison of timely data transfer rate for the networks with different sizes and node densities.

## VIII. CONCLUSION

In this paper, we present an evolutionary game based secure clustering protocol with fuzzy trust evaluation and outlier detection which is called EGSCFO to assure secured communication in WSNs. Our protocol is fully distributed that each sensor node independently estimates the trust values of other nodes using an IT2 fuzzy logic based trust evaluation method which can effectively deal with the trust uncertainty problem. Combined with trust recommendation mechanism, each sensor node updates the trust values which are further analyzed via a K-Means based outlier detection scheme to detect the malicious. To isolate malicious nodes from being cluster heads, each sensor node competes for the cluster head by playing an evolutionary game with its adjacent trusted nodes while a trade-off between security assurance and energy saving can be acquired. A node failed to be the cluster head chooses a trusted cluster head to join in the corresponding cluster. Simulation results verify that our protocol EGSCFO can effectively defend the network against the attacks from internal selfish or compromised nodes, since the number of malicious clusters where the malicious nodes act as cluster heads rapidly decreases with fluctuations until reaches the minimal value 0. Then our protocol can obviously enhance the data delivery rate and reduce the transmission delay due to the malicious nodes are successfully isolated from transmission. In future work, we plan to study how to establish standard misbehavior or malicious transmission patterns which can be used to distinguish the malicious nodes from normal ones according to their transmission behaviors.

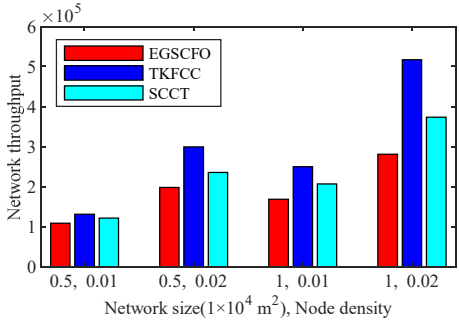

Fig. 16. The comparison of network throughput for the networks with different sizes and node densities.

because the malicious nodes have more neighbor nodes so that more chance can be acquired to launch the attacks.

Fig. 16 gives the comparison of network throughput for the networks with different sizes and node densities. It shows that for the networks with the same size, the one with bigger node density is with higher throughput for all these protocols since more sensor nodes are deployed in the network. In addition, for the networks with the same number of sensor nodes, the one with larger size has lower throughput. This is because sensor nodes in the network with larger size have bigger average distance to base station, so that the network lifetime is shortened that results in the reduction of network throughput.

The comparison of timely data transfer rate for the networks with different sizes and node densities is shown in Fig. 17. From this figure, we can see that for all the instances of the networks with different sizes and node densities, our protocol EGSCFO has the highest timely data transfer rate, followed by SCCT, and the final is TKFCC. This situation is consistent with the performance of protecting the network from packet dropping and delaying attacks. In addition, this figure also shows that for the networks with a smaller size, the one with bigger node density has lower timely data transfer rate for all these protocols due to more attacks may occur in the network. For the networks with a larger size, the above conclusion is still applicable for EGSCFO while the converse is true for TKFCC and SCCT. This is because for the network with a larger size, with the increase of node density, some increments of network throughput can offset the negative impacts of the increased attacks on the timely data transfer rate.

## REFERENCES

[1] L. Yang, Y. Z. Lu, Y. C. Zhong, and S. X. Yang, "An unequal cluster-based routing scheme for multi-level heterogeneous wireless sensor networks," *Telecommun. Syst.*, vol. 68, no. 1, pp. 11-26, May 2018.
[2] M. Ayaz, M. Ammad-uddin, I. Baig, and E. M. Aggoune, "Wireless Sensor's Civil Applications, Prototypes, and Future Integration Possibilities: A Review," *IEEE Sensors J.*, vol. 18, no. 1, pp. 4-30, Jan. 2018.
[3] I. Tomic, and J. A. McCann, "A Survey of Potential Security Issues in Existing Wireless Sensor Network Protocols," *IEEE Internet Things J.*, vol. 4, no. 6, pp. 1910-1923, Dec. 2017.
[4] D. P. Kumar, T. Amgoth, and C. S. R. Annavarapu, "Machine learning algorithms for wireless sensor networks: A survey," *Inf. Fusion*, vol. 49, pp. 1-25, Sep. 2019.

[5] G. S. Yang, T. T. Liang, X. Y. He, and N. X. Xiong, "Global and Local Reliability-Based Routing Protocol for Wireless Sensor Networks," *IEEE Internet Things J.*, vol. 6, no. 2, pp. 3620-3632, Apr. 2019.

[6] M. Pavani and P. Trinatha Rao, "Adaptive PSO with optimised firefly algorithms for secure cluster-based routing in wireless sensor networks," *IET Wirel. Sens. Syst.*, vol. 9, iss. 5, pp. 274-283, Jun. 2019.

[7] R. Wang, Z. Y. Zhang, Z. W. Zhang, and Z. P. Jia, "ETMRM: An Energy-efficient Trust Management and Routing Mechanism for SDWSNs," *Comput. Netw.*, vol. 139, pp. 119-135, Jul. 2018.

[8] W. A. Aliady, and S. A. Al-Ahmadi, "Energy Preserving Secure Measure Against Wormhole Attack in Wireless Sensor Networks," *IEEE Access*, vol. 7, pp. 84132-84141, 2019.

[9] H. Jadidoleslamy, M. R. Aref, and H. Bahramgiri, "A fuzzy fully distributed trust management system in wireless sensor networks," *Int. J. Electron. Commun.*, vol. 70, no. 1, pp. 40-49, 2016.

[10] A. Ahmed, K. Abu Bakar, M. I. Channa, K. Haseeb, and A. W. Khan, "TERP: A Trust and Energy Aware Routing Protocol for Wireless Sensor Network," *IEEE Sensors J.*, vol. 15, no. 12, pp. 6962-6972, Dec. 2015.

[11] B. A. Ali, H. M. Abdulsalam, and A. AlGhemlas, "Trust Based Scheme for IoT Enabled Wireless Sensor Networks," *Wireless Pers. Commun.*, vol. 99, no. 2, pp. 1061-1080, Mar. 2018.

[12] A. Saidi, K. Benahmed, and N. Seddiki, "Secure cluster head election algorithm and misbehavior detection approach based on trust management technique for clustered wireless sensor networks," *Ad Hoc Netw.*, vol. 106, Sep. 2020.

[13] M. Udhayavani, and M. Chandrasekaran, "Design of TAREEN (trust aware routing with energy efficient network) and enactment of TARF: a trust-aware routing framework for wireless sensor networks," *Cluster Comput.*, vol. 22, pp. 11919-11927, Sep. 2019.

[14] R. A. Shaikh, and A. S. Alzahrani, "Trust Management Method for Vehicular Ad Hoc Networks," in *Int. Conf. Heterogeneous Netw.*, Berlin, Germany, 2013, pp. 801-815.

[15] J. S. Fu, and Y. Liu, "Double Cluster Heads Model for Secure and Accurate Data Fusion in Wireless Sensor Networks," *Sensors*, vol. 15, no. 1, pp. 2021-2040, Jan. 2015.

[16] L. Yang, Y. Z. Lu, S. Liu, T. Guo, and Z. F. Liang, "A Dynamic Behavior Monitoring Game-Based Trust Evaluation Scheme for Clustering in Wireless Sensor Networks," *IEEE Access*, vol. 6, pp. 71404-71412, 2018.

[17] A. Ahmed, K. Abu Bakar, M. I. Channa, and A. W. Khan, "A Secure Routing Protocol with Trust and Energy Awareness for Wireless Sensor Network," *Mobile Netw. Appl.*, vol. 21, no. 2, pp. 272-285, Apr. 2016.

[18] Y. X. Liu, M. X. Dong, K. Ota, and A. F. Liu, "ActiveTrust: Secure and Trustable Routing in Wireless Sensor Networks," *IEEE Trans. Inf. Forensics Secur.*, vol. 11, no. 9, pp. 2013-2027, Sep. 2016.

[19] R. W. Anwar, A. Zainal, F. Outay, A. Yasar, and S. Iqbal, "BTEM: Belief based trust evaluation mechanism for Wireless Sensor Networks," *Future Gener. Comput. Syst.*, vol. 96, pp. 605-616, Jul. 2019.

[20] M. A. Javed, S. Zeadally, and Z. Hamid, "Trust-based security adaptation mechanism for Vehicular Sensor Networks," *Comput. Netw.*, vol. 137, pp. 27-36, Jun. 2018.

[21] W. Fang, W. Zhang, W. Chen, Y. Liu, and C. Tang, "TMSRS: trust management-based secure routing scheme in industrial wireless sensor network with fog computing," *Wireless Netw.*, vol. 26, no. 5, pp. 3169-3182, Jul. 2020.

[22] G. J. Han, Y. He, J. F. Jiang, N. Wang, M. Guizani, and J. A. Ansere, "A Synergetic Trust Model Based on SVM in Underwater Acoustic Sensor Networks," *IEEE Trans. Veh. Technol.*, vol. 68, no. 11, pp. 11239-11247, Nov. 2019.

[23] T. Zhang, L. S. Yan, and Y. Yang, "Trust evaluation method for clustered wireless sensor networks based on cloud model," *Wireless Netw.*, vol. 24, no. 3, pp. 777-797, Apr. 2018.

[24] S. Augustine, and J. P. Ananth, "Taylor kernel fuzzy C-means clustering algorithm for trust and energy-aware cluster head selection in wireless sensor networks," *Wireless Netw.*, Jun. 2020.

[25] P. Rodrigues, and J. John, "Joint trust: an approach for trust-aware routing in WSN," *Wireless Netw.*, vol. 26, no. 5, pp. 3553-3568, Jul. 2020.

[26] T. Kalidoss, L. Rajasekaran, K. Kanagasabai, G. Sannasi, and A. Kannan, "QoS Aware Trust Based Routing Algorithm for Wireless Sensor Networks," *Wireless Pers. Commun.*, vol. 110, no. 4, pp. 1637-1658, Feb. 2020.

[27] K. Thangaramya, K. Kulothungan, S. I. Gandhi, M. Selvi, S. Kumar, and K. Arputharaj, "Intelligent fuzzy rule-based approach with outlier detection for secured routing in WSN," *Soft Comput.*, Apr. 2020.

[28] L. Yang, Y. Z. Lu, Y. C. Zhong, X. G. Wu, and S. J. Xing, "A hybrid, game theory based, and distributed clustering protocol for wireless sensor networks," *Wireless Netw.*, vol. 22, no. 3, pp. 1007-1021, Apr. 2016.

[29] W. H. R. Chan, P. F. Zhang, I. Nevat, S. G. Nagarajan, A. C. Valera, H. X. Tan, and N. Gautam, "Adaptive Duty Cycling in Sensor Networks With Energy Harvesting Using Continuous-Time Markov Chain and Fluid Models," *IEEE J. Sel. Areas Commun.*, vol. 33, no. 12, pp. 2687-2700, Dec. 2015.

[30] S. Tan, X. Li, and Q. Dong, "A Trust Management System for Securing Data Plane of Ad-Hoc Networks," *IEEE Trans. Veh. Technol.*, vol. 65, no. 9, pp. 7579-7592, Sep. 2016.

[31] W. Peng, C. D. Li, G. Q. Zhang, and J. Q. Yi, "Interval type-2 fuzzy logic based transmission power allocation strategy for lifetime maximization of WSNs," *Eng. Appl. Artif. Intell.*, vol. 87, pp. 11, Jan. 2020.

[32] D. R. Wu, and J. M. Mendel, "Recommendations on designing practical interval type-2 fuzzy systems," *Eng. Appl. Artif. Intell.*, vol. 85, pp. 182-193, Oct. 2019.

[33] D. R. Wu, "On the Fundamental Differences Between Interval Type-2 and Type-1 Fuzzy Logic Controllers," *IEEE Trans. Fuzzy Syst.*, vol. 20, no. 5, pp. 832-848, Oct. 2012.

[34] L. Yang, Y. Z. Lu, L. Xiong, Y. Tao, and Y. C. Zhong, "A Game Theoretic Approach for Balancing Energy Consumption in Clustered Wireless Sensor Networks," *Sensors*, vol. 17, no. 11, Nov. 2017.

[35] K. A. Gemeda, G. Gianini, and M. Libsie, "An evolutionary cluster-game approach for Wireless Sensor Networks in non-collaborative settings," *Pervasive Mob. Comput.*, vol. 42, pp. 209-225, Dec. 2017.

[36] Z. H. Tian, X. S. Gao, S. Su, J. Qiu, X. J. Du, and M. Guizani, "Evaluating Reputation Management Schemes of Internet of Vehicles Based on Evolutionary Game Theory," *IEEE Trans. Veh. Technol.*, vol. 68, no. 6, pp. 5971-5980, Jun. 2019.

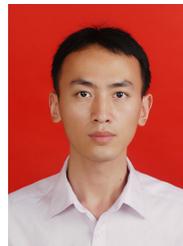

**Liu Yang** received his B.S. degree in Electronic Information Science and Technology from Qingdao University of Technology, Shandong, China, in 2010, and Ph.D. degree in Communication and Information Systems at the College of Communication Engineering, Chongqing University, Chongqing, China, in 2016. He is now a lecturer in Chongqing University of Posts and Telecommunications. His research interests include Internet of Things, data analysis, and artificial intelligence.

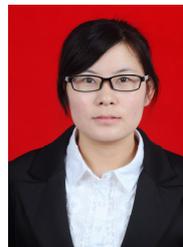

**Yinzhi Lu** received her M.S. degree in Communication and Information Systems from Chongqing University, Chongqing, China, in 2014. She is currently pursuing the Ph. D. degree in Information and Communication Engineering with the School of Communication and Information Engineering, Chongqing University of Posts and Telecommunications, Chongqing, China. She was a teaching assistant with the School of Electronic Information Engineering, Yangtze Normal University from 2014 to 2019. Her current research interests include Internet of Things, time sensitive network, and artificial intelligence.

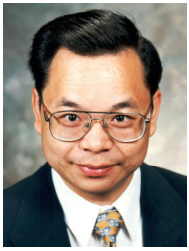
**Simon X. Yang** (S'97–M'99–SM'08) received the B.Sc. degree in engineering physics from Beijing University, Beijing, China, in 1987, the first of two M.Sc. degrees in biophysics from the Chinese Academy of Sciences, Beijing, China, in 1990, the second M.Sc. degree in electrical engineering from the University of Houston, Houston, TX, in 1996, and the Ph.D. degree in electrical and computer engineering from the University of Alberta, Edmonton, AB, Canada, in 1999.

Dr. Yang is currently a Professor and the Head of the Advanced Robotics and Intelligent Systems Laboratory at the University of Guelph, Guelph, ON, Canada. His research interests include robotics, intelligent systems, sensors and multi-sensor fusion, wireless sensor networks, control systems, machine learning, fuzzy systems, and computational neuroscience.

Prof. Yang has been very active in professional activities. He serves as the Editor-in-Chief of *International Journal of Robotics and Automation*, and an Associate Editor of *IEEE Transactions on Cybernetics*, *IEEE Transactions on Artificial Intelligence*, and several other journals. He has involved in the organization of many international conferences.

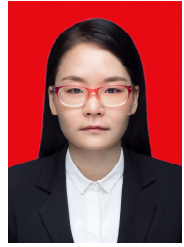
**Zhifang Liang** received her B.S. degree in Computer Science and Technology from Shanxi Normal University, Linfen, China, in 2013, and received her Ph.D. degree in intelligent signal processing from the School of Communication Engineering, Chongqing University, Chongqing, China, in 2017. Currently she works in the school of Communication and Information Engineering, Chongqing University of Posts and Telecommunications. Her research interests include electronic nose technology, Internet of Things, and machine learning.

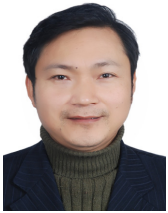
**Yuanchang Zhong** received the B.S. degree in Changchun University of Science and Technology, Changchun, China, in 1988, and the M.S. and Ph.D. degrees from Chongqing University, Chongqing, China, in 2004 and 2012. From 2003 to 2011, he was an Assistant Professor with the College of Communication Engineering, Chongqing University. He was a Professor with the College of Communication Engineering, Chongqing University from 2012 to 2020. He is currently a Professor at the School of Electrical Engineering, Chongqing University. His research interests include wireless power transfer, Internet of Things, and artificial intelligence.

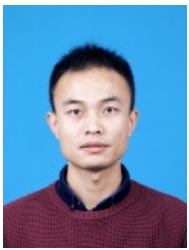
**Tan Guo** received his M.S. degree in Signal and Information Processing from Chongqing University, Chongqing, China, in 2014, and Ph.D. degree in Communication and Information Systems from Chongqing University, Chongqing, China, in 2017. He is now a lecturer in Chongqing University of Posts and Telecommunications. His research interests include Internet of Things, biometrics, pattern recognition, and machine learning.